\documentclass[conference,a4paper]{IEEEtran}

\usepackage{amsmath,amsfonts,color,comment,cite}
\usepackage[pdftex]{graphicx}

\newcommand{\ee}{\mathrm{e}}

\newcommand{\K}{\mathcal{K}}

\newtheorem{theorem}{Theorem}
\newtheorem{lemma}[theorem]{Lemma}
\newtheorem{definition}[theorem]{Definition}

\begin{document}

\title{On the Optimality of Some\\Group Testing Algorithms}

\author{\IEEEauthorblockN{Matthew Aldridge}
\IEEEauthorblockA{Department of Mathematical Sciences\\
University of Bath, Bath, U.K.\\
Heilbronn Institute for Mathematical Research\\
Bristol, U.K. \\
Email: m.aldridge@bath.ac.uk}}

\maketitle

\begin{abstract}
We consider Bernoulli nonadaptive group testing with $k = \Theta(n^\theta)$ defectives, for $\theta \in (0,1)$. The practical definite defectives (DD) detection algorithm is known to be optimal for $\theta \geq 1/2$. We give a new upper bound on the rate of DD, showing that DD is strictly suboptimal for $\theta < 0.41$. We also show that the SCOMP algorithm and algorithms based on linear programming achieve a rate at least as high as DD, so in particular are also optimal for $\theta \geq 1/2$.
\end{abstract}

\section{Introduction}

Group testing is a sparse inference problem that involves detecting the defective items in a population through the means of pooled tests. By testing a `pool' -- that is, a subset -- of items, we discover whether that pool is free of defective items or whether it contains at least one defective item. After a sequence of such tests, we must infer (or `detect') the true set of defective items. The goal is to find the defective set with low probability of error using as few tests as possible.

This naturally splits into two problems: first, designing the testing strategy, and second, detecting the defective items, given the test design and the results of the test.

For the design, here we consider Bernoulli nonadaptive testing. In nonadaptive testing, the test pools are all defined in advance, so the tests can be carried out in parallel. Bernoulli nonadaptive testing is the simplest form of nonadaptive testing, where each item is placed in each test independently with some fixed probability $p$. Bernoulli nonadaptive testing is simple, order optimal for all $\theta \in [0,1)$ (see \cite{chan,ABJ,ABG,sc}), even has optimal constants for $\theta \in [0,1/3)$ (see \cite{sc,sc2,aldridge}), and has been studied by many authors (see \cite{chan,chan2,ABJ,ABG,sc,sc2,aldridge,malyutov,sebo,mal,laarhoven} for just a few examples).

For the detection problem, we consider three detection algorithms: one called DD (for `definite defectives'), one called SCOMP (which can be seen as a Sequential version of the earlier COMP algorithm), and approaches based on linear programming. In contrast to the theoretical algorithms often used to prove achievability results, these three algorithms are practical, in that they are computationally efficient and do not require knowledge of the true number of defectives. It is important to know when these practical algorithms are as good as the theoretically optimal but impractical detection algorithms. Here, we study this question in the context of Bernoulli nonadaptive testing. It is known that DD is optimal for denser cases, where we have more defectives.  Our first result here is that DD is strictly suboptimal for sparser cases. Our second result is that SCOMP and linear programming approaches are also optimal in denser cases.

We begin with some notation. There are $n$ items, of which $k$ are defective, and we use $T$ tests. We assume that defectivity is rare, in that $k = o(n)$, and concentrate on the case where $k$ scales like $k = \Theta(n^\theta)$ for some sparsity parameter $\theta \in (0,1)$. (The $\theta = 0$ case behaves much the same, but one has to be a little careful over limiting arguments, as $k$ does not tend to $\infty$ as $n$ grows. We omit this case for brevity, although the results of this paper do carry over.)
We assume that the true defective set $\K$ is chosen uniformly at random for all subsets of size $k$.

A useful way to keep track of the design of the test pools is to write $x_{ti} = 1$ to denote that item $i \in \{1,2,\dots,n\}$ is in the pool for test $t \in \{1,2,\dots,T\}$, and $x_{ti} = 0$ to denote that item $i$ is not in the pool for test $t$.

\begin{definition}
In \emph{Bernoulli nonadaptive group testing} we set $x_{ti} = 1$ with probability $p$ and $x_{ti} = 0$ with probability $1-p$, independently over $t \in \{1,2,\dots,T\}$ and $i \in \{1,2,\dots,n\}$, for some fixed $p \in [0,1]$.
\end{definition}

Recall that in standard group testing, a test outcome is positive if at least one item from $\K$ is in the test, and negative if no items from $\K$ are in the test. That is, writing $y_t$ for the outcome of test $t$, we have
  \[ \label{eq:outcome}
    y_t = \begin{cases} 1 & \text{if there exists $i \in \K$ with $x_{ti} = 1$,} \\
             0 & \text{if for all $i \in \K$ we have $x_{ti} = 0$.} \end{cases}
  \]

In general, since there are $\binom nk$ possible defective sets and $2^T$ possible outcomes from $T$ tests, we see that we have a lower bound on the number of tests required to reconstruct the defective set of
  \[ T \geq \log_2 \binom nk \sim k \log_2 \frac nk  \sim (1 - \theta) k \log_2 n. \]
One way of interpreting this is to say that specifying the defective set requires $\log_2 \binom nk$ bits of information, and each test, with a yes--no answer, can impart at most $1$ bit of information. (From now on, $\log$ always means $\log_2$.) Given this, for a scheme that uses $T$ tests, we can regard the `bits learned per test' $\log \binom nk/T$
as the \emph{rate} of group testing. (This terminology follows that of Baldassini, Johnson, and Aldridge \cite{BJA}.)

\begin{definition}
The \emph{average error probability} is defined as follows: if our estimate of the defective set $\K$ is $\hat\K$, then the error probability is
  \[ \frac{1}{\binom nk} \sum_{\substack{\K \subset \{1, \dots, n\} \\ |\K| = k}} \mathbb P(\hat\K \neq \K) , \]
where the probability is over the random choice of test design.
\end{definition}

\begin{definition}
Given a detection algorithm, we say a rate $R$ is \emph{achievable} if, for any $\epsilon > 0$, then for $n$ sufficiently large and $k = k(n) = \Theta(n^\theta)$, there exists a Bernoulli parameter $p = p(n)$ and a number of tests $T = T(n)$ with rate
\[ \frac{ \log \binom nk}{T} > R \]
and Bernoulli nonadaptive testing having error probability less than $\epsilon$.
\end{definition}

The capacity of Bernoulli nonadaptive group testing (that is, the maximum achievable rate for any detection algorithm) is
  \begin{equation} \label{uni}
    C = \max_{\nu > 0} \min \left\{ h(\mathrm e^{-\nu}),\, \frac{\nu}{\ee^\nu \ln 2} \frac{1-\theta}{\theta} \right\} .
  \end{equation}
Achievability was shown by Scarlett and Cevher \cite{sc} and the converse by Aldridge \cite{aldridge}. It is easy to check that for most $\theta$ this can be simplified to
  \begin{equation*}
    C = \begin{cases} 1 & \text{for }\theta \leq \tfrac13 ,\\
          0.531 \displaystyle\frac{1-\theta}{\theta} & \text{for }\theta > 0.359 ,
        \end{cases}
  \end{equation*}
so the $\max$-$\min$ is only needed in the small interval $\theta \in (1/3,0.359)$.
  
However, the achievability result of Scarlett and Cevher \cite{sc} does not give a practical decoding algorithm to achieve the rate \eqref{uni}.

Aldridge, Baldassini and Johnson \cite{ABJ} showed that a simple algorithm called DD (for `definite defectives') is optimal for $\theta \geq 1/2$. We define the DD algorithm in Section II. Also, here `optimal' is used throughout to mean `achieves the capacity \eqref{uni}.' One might wonder if DD is also optimal for $\theta < 1/2$. Our first main result mostly answers this in the negative: we give a new upper bound for the rate of DD, showing that for $\theta < 0.407$ DD is strictly suboptimal (Theorem \ref{ddthm}).

\begin{figure}
\begin{center}
\includegraphics[width=0.48\textwidth]{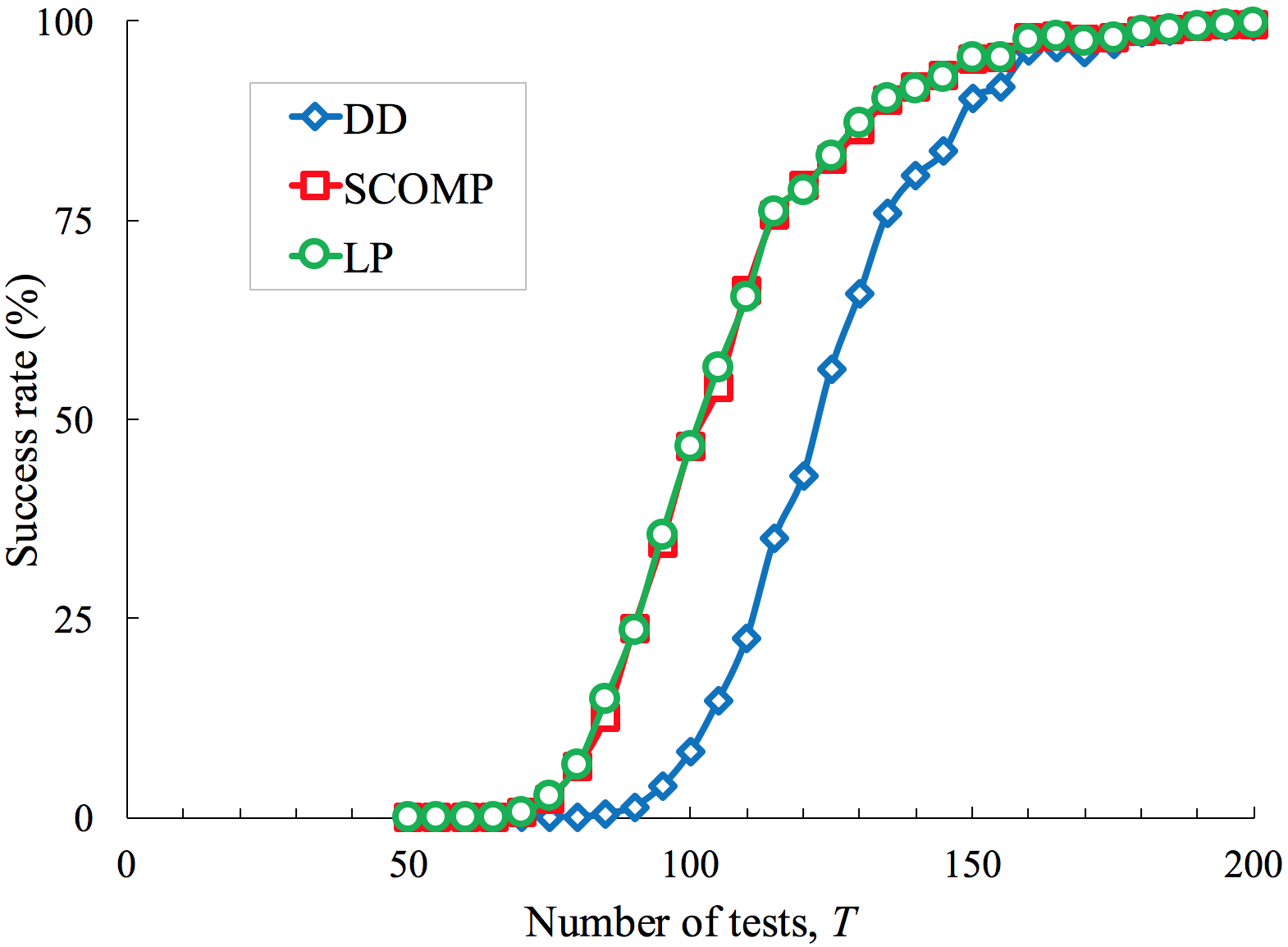}
\end{center}
\caption{Performance of Bernoulli nonadaptive group testing algorithms. The parameters were $n = 500$, $k = 10$, and $p = 1/(k+1) \approx 0.0909$. Linear programming uses the `Malioutov rule' $\hat{\mathcal K} = \{i : z_i > 0 \}$ (see Subsection \ref{sec:lp}). Note that performance of  LP and SCOMP is so similar that the SCOMP line can be difficult to see. Each point represents $1000$ simulated runs.}\label{fig:sim}
\end{figure}

Two decoding algorithms seem to perform better than DD in practice -- thus far without theoretical justification. These are the SCOMP algorithm of Aldridge, Baldassini, and Johnson \cite{ABJ}, and approaches based on linear programming (see, for example, \cite{mal,chan2}). We define these algorithms in Section III-B and III-C. Figure \ref{fig:sim} shows a simulation of all three algorithms, demonstrating the superiority of SCOMP and linear programming. This theoretical performance suggests that the rate of these algorithms should be at least as high as DD, so in particular they should also be optimal for $\theta > 1/2$. Our second main result proves this (Theorems \ref{scompthm} and \ref{lpthm}).

\begin{figure}
\begin{center}
\includegraphics[width=0.48\textwidth]{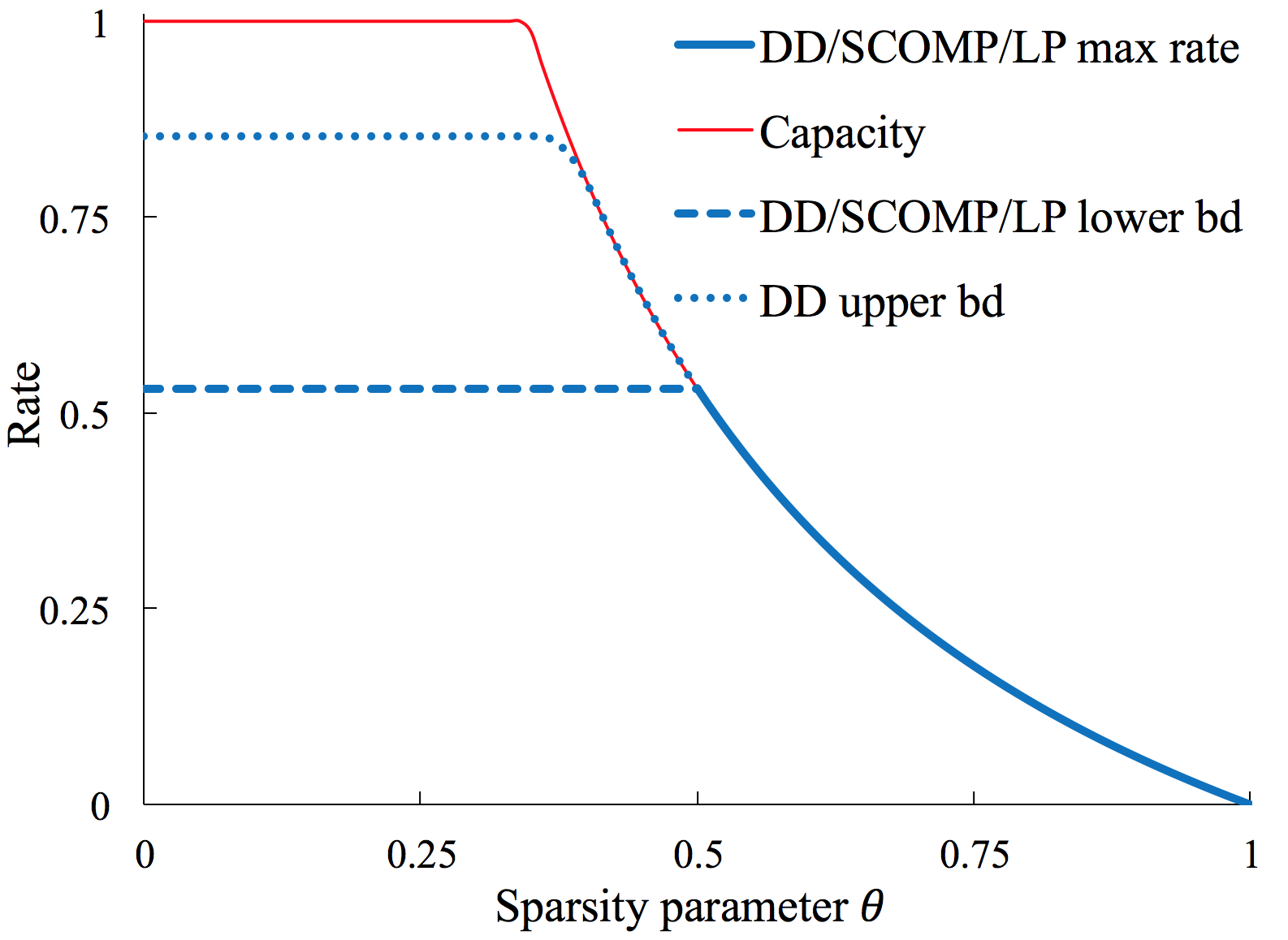}
\end{center}
\caption{Graph showing rates for Bernoulli nonadaptive group testing: capacity \eqref{uni}; maximum achievable rate for DD, SCOMP and LP (\eqref{exact}, Theorem \ref{scompthm} and  Theorem \ref{lpthm}); lower bound for DD, SCOMP and LP (\eqref{ddbd}, Theorem \ref{scompthm} and  Theorem \ref{lpthm}); and upper bound for DD (Theorem \ref{ddthm}).}\label{DDgraph}
\end{figure}

The rate bounds implied by our main results are illustrated in Figure \ref{DDgraph}.

\section{DD is suboptimal for $\theta < 0.42$}

The DD algorithm of Aldridge, Baldassini, and Johnson \cite{ABJ} works as follows:
\begin{enumerate}
  \item Any item in a negative test is definitely nondefective. The remaining items we call `possible defectives'.
  \item If a (necessarily positive) test contains exactly one possible defective, then that item is in fact definitely defective (`DD').
  \item The definitely defective items are declared to be defective. All other items -- the definitely nondefective items and any remaining possible defectives -- are declared to be nondefective.
\end{enumerate}

DD is `practical' in the sense that it requires only $O(nT) = O(kn \log n) = O(n^{1 + \theta}\log n)$ work and storage -- which is what is required just to store and read the test design. In comparison, the decoder of Scarlett and Cevher \cite{sc} involves solving the maximum likelihood problem, which is in NP in general. Further, DD does not require the detector to know $k$ exactly or even approximately, while the maximum likelihood decoder requires the exact value of $k$.

It was shown in \cite{ABJ} that, with Bernoulli tests, the DD algorithm can achieve a rate of
  \begin{equation} \label{ddbd}
    R^*_{\mathrm{DD}} \geq \frac{1}{\ee \ln 2}\min \left\{ 1, \frac{1 - \theta}{\theta} \right\} ,
  \end{equation}
where $1/\ee\ln2 \approx 0.531$.
  
This can be compared with the capacity for Bernoulli testing \eqref{uni}. Since \eqref{uni} gives an upper bound of $(1/\ee \ln 2) (1-\theta)/\theta$ for $\theta > 0.359$, and \eqref{ddbd} gives the same rate for $\theta \geq 1/2$, we see that for $\theta \geq 1/2$ we have the equality
  \begin{equation} \label{exact}
    R^*_{\mathrm{DD}} = \frac{1}{\ee \ln 2} \frac{1- \theta}{\theta} = 0.531 \frac{1- \theta}{\theta}  \qquad \text{for } \theta \geq \frac12 ,
  \end{equation}
and DD is optimal in this region. (This was first noted in \cite{aldridge}.)

Here we give an upper bound for DD, which shows that it is strictly suboptimal for some smaller values of $\theta$.

\begin{theorem} \label{ddthm}
The rate of DD for Bernoulli nonadaptive group testing is bounded above by 
  \begin{equation} \begin{split} \label{newbd}
    \hspace{-1.3cm} R^*_{\mathrm{DD}} \leq \max_{\nu > 0} \min \bigg\{ \ee^{-\nu} \log \ee^\nu + \nu\ee^{-\nu} \log \frac{1}{1 - \ee^{-\nu}} , \\ 
  \frac{\nu}{\ee^\nu \ln 2}  \frac{1-\theta}{\theta} \bigg \} . \hspace{-0.8cm}
  \end{split}\end{equation}
In particular, DD is strictly suboptimal for $\theta < \theta^*$, where
  \begin{equation*}
    \theta^* = \frac{1}{2-\ln (1 - \ee^{-1})} = 0.407. 
  \end{equation*}
\end{theorem}

Although the bound \eqref{newbd} appears complicated, it is easy to check that for $\theta > \theta^*$ the second minimand dominates and gives the same bound $0.531 (1- \theta)/\theta$ as the standard capacity bound \eqref{uni}. Further, one can check numerically that for $\theta < 0.357$ 
the first minimand of \eqref{newbd} dominates and gives a bound of $0.853$. 
It is only for $\theta \in (0.357,0.407)$ that the complicated expression \eqref{newbd} is necessary. Whether or not DD is optimal in the range $\theta \in (0.407,1/2)$ remains an open question.

The above bounds are illustrated in Figure \ref{DDgraph}.



\begin{IEEEproof}
The key point is that DD can make use of tests with no defectives (to eliminate `possible defectives'), and tests with one defective (if there are no other possible defectives in the test, the defective is `definitely defective'), but does not make use of any tests containing two or more defectives (they necessarily contain at least two possible defectives).

Suppose we use independent tests (as we do with a Bernoulli design). Write $q_d$ for the probability a test contains exactly $d$ defectives and $q_{d+}$ for the probability a test contains at least $d$ defectives. Then a standard information argument gives a bound in terms of the entropy
  \begin{align*}
    R &\leq h(q_0) \\
      &= q_0 \log \frac{1}{q_0} + q_{1+} \log \frac{1}{q_{1+}} \\
      &= q_0 \log \frac{1}{q_0} + (q_1 + q_{2+}) \log \frac{1}{q_{1+}}. \end{align*}
(Note that when $q_0 = q_{1+} = 1/2$, this gives the standard counting bound $R \leq 1$.) While it is helpful to think of this in terms of `bits learned per tests', it can be made rigorous using the concept of typical sets -- see, for example, \cite[Lemma 1]{aldridge} -- which tells us that with high probability we see one of $2^{Th(q_0)}$ `typical' outcomes.

Since DD does make use of those tests containing $2$ or more defectives, we get a tighter bound
  \begin{equation} \label{entbd}
    R^*_{\mathrm{DD}} \leq q_0 \log \frac{1}{q_0} + q_1 \log \frac{1}{q_{1+}}.
  \end{equation}
This is because we learn at most $\log 1/q_0$ bits with probability $q_0$ from the negative tests, and from positive tests we learn $\log 1/q_{1+}$ bits only with probability $q_{1}$, since DD gets no information from a (positive) test with probability $q_{2+}$.

Now recall we are using a Bernoulli test design. Setting the parameter as $p \sim \nu / k$ (as we must to get a nonzero rate \cite{aldridge}), we have
  \begin{gather*}
    q_0 = (1 - p)^k \sim \left(1 - \frac{\nu}{k}\right)^k \to \ee^{-\nu} ,\\
    q_1 = kp(1-p)^{k-1} \sim k \frac{\nu}{k}\left(1 - \frac{\nu}{k}\right)^{k-1} \to \nu \ee^{-\nu} , \\
    q_{1+} = 1 - q_0\sim 1 - \ee^{-\nu} .
  \end{gather*}
as $k \to \infty$. Hence the bound \eqref{entbd} becomes
  \[
    R^*_{\mathrm{DD}} \leq \ee^{-\nu} \log \ee^\nu + \nu\ee^{-\nu} \log \frac{1}{1 - \ee^{-\nu}} . \]

Combined with the capacity bound \eqref{uni}, this gives
  \begin{multline*} R^*_{\mathrm{DD}} \leq \max_{\nu > 0} \min \bigg\{ \ee^{-\nu} \log \ee^\nu + \nu\ee^{-\nu} \log \frac{1}{1 - \ee^{-\nu}} , \\ h(\mathrm e^{-\nu})\,, \  
  \frac{\nu}{\ee^\nu \ln 2}  \frac{1-\theta}{\theta} \bigg \} . \end{multline*}
Note, however, that the second minimand 
  \[ h(\ee^{-\nu}) = q_0 \log \frac{1}{q_0} + q_{1+} \log \frac{1}{q_{1+}} \]
is always dominated by the first minimand, so the second minimand can be deleted.

For the final part of the theorem, we note that second minimand in \eqref{newbd} is maximised at $\nu = 1$, and this dominates until
  \[ \ee^{-1} \log \ee + \ee^{-1} \log \frac{1}{1 - \ee^{-1}} < \frac{1}{\ee\ln2} \frac{1-\theta}{\theta} . \]
By rearranging, and noting that $\log \ee = 1/\ln 2$, we see that this is precisely when $\theta < \theta^*$.
\end{IEEEproof}

We remark that a similar proof technique could be used to bound the rate of the COMP algorithm of Chan et al \cite{chan}. This algorithm makes no use of of any positive tests, so the equivalent to the bound \eqref{entbd} would be
  \[ R^*_{\mathrm{COMP}} \leq q_0 \log \frac{1}{q_0}. \]
Taking $q_0 \sim \ee^{-\nu}$ as above, and optimising at $\nu = 1$ gives the bound
  \[ R^*_{\mathrm{COMP}} \leq \ee^{-1} \log \ee = \frac{1}{\ee \ln 2} . \]
Compared to the actual maximum achievable rate of COMP \cite{chan,aldridge}, which is
  \[ R^*_{\mathrm{COMP}} = \frac{1}{\ee \ln 2} (1 - \theta) , \]
we see that such a bound is tight for $\theta = 0$ but loose for $\theta > 0$.


\section{Algorithms that are optimal for $\theta \geq 1/2$}

We have seen that the DD algorithm with Bernoulli testing achieves the rate \eqref{ddbd}. Together with the capacity bound \eqref{uni}, we see that DD is optimal for Bernoulli testing for $\theta \geq 1/2$.

As we saw in Figure \ref{fig:sim}, the SCOMP algorithm \cite{ABJ} and approaches based on linear programming \cite{mal,chan2} outperform DD in simulations. In this section, we prove that the SCOMP algorithm and LP approaches achieve at least as high a rate as DD, so in particular satisfy the bound \eqref{ddbd} and are also optimal for $\theta \geq 1/2$.

\subsection{Property $P$}

The crucial step to showing an algorithm outperforms DD is the a property we call `Property $P$'. First recall the following definition.

\begin{definition}
A set $\mathcal L \subset \{1,2,\dots,n\}$ is \emph{satisfying} if  no negative test contains any item from $\mathcal L$ and every positive test contains at least one item from $\mathcal L$.
\end{definition}

So a satisfying set is one that `fits' the observations. In particular, the true defective set $\K$ is always a satisfying set, although their may be others.

\begin{definition}
A detection algorithm \texttt{A} has  \emph{Property $P$} if, whenever the DD algorithm outputs a satisfying set, then \texttt{A} also outputs that same satisfying set. 
\end{definition}

\begin{lemma} \label{lem}
Suppose the detection algorithm \texttt{A} satisfies Property $P$. Then \texttt{A} succeeds whenever DD succeeds. In particular, with Bernoulli nonadaptive group testing, we have
  \[ R^*_{\text{\texttt{A}}} \geq R^*_{\mathrm{DD}} \geq\min \left\{ \frac{1}{\ee \ln 2},\, \frac{1}{\ee \ln 2} \frac{1 - \theta}{\theta} \right\} , \]
and $\texttt{A}$ is optimal for $\theta \geq 1/2$.
\end{lemma}

\begin{IEEEproof}
Note that DD is not guaranteed to output a satisfying set. If the output of DD is not satisfying, then clearly the algorithm is in error. If the output of DD is satisfying, then it is necessarily the unique smallest satisfying set (since each definite defective appears uniquely in some test, so must be in the true defective set), and the algorithm may be successful.

Since algorithm \texttt{A} has Property $P$, in this latter case, \texttt{A} also outputs this unique smallest satisfying set. Hence, whenever DD succeeds, \texttt{A} will succeed also.

The second part of the lemma then follows immediately from \eqref{ddbd}.
\end{IEEEproof}

\subsection{SCOMP}

SCOMP is an algorithm due to Aldridge, Baldassini, and Johnson \cite{ABJ} that builds a satisfying set by starting from the DD set of definite defectives and sequentially adding new items until a satisfying set is reached. (The name comes from `Sequential COMP', as it can be viewed as a sequential version of the COMP algorithm of Chan et al \cite{chan}.)

Although recalling the precise details of how SCOMP extends from DD to a satisfying set is not required for the proof of following result, we define it here for completeness.
\begin{enumerate}
  \item Any item in a negative test is definitely nondefective. The remaining items we call `possible defectives'.
  \item If a (necessarily positive) test contains exactly one possible defective, then that item is in fact definitely defective (`DD').
  \item The definitely defective items are declared to be defective, we call these $\hat\K$, and definitely nondefective items are declared to be nondefective. The other possible defectives are not yet declared either way.
  \item Any positive test is called `unexplained' if it does not contain any items from $\hat\K$. Add to $\hat\K$ the possible defective not in $\hat\K$ that is in the most unexplained tests, marking the corresponding tests as no longer unexplained. (Ties may be broken arbitrarily.)
  \item Repeat step 4 until no tests remain unexplained. The estimate of the defective set is $\hat\K$.
\end{enumerate}

Note that a satisfying set leaves no unexplained tests, and that any set containing no definite nondefectives and leaving no unexplained tests is satisfying. Note also that the set of all possible defectives is satisfying, so the SCOMP algorithm does indeed terminate.

\begin{theorem} \label{scompthm}
SCOMP with Bernoulli tests can achieve the rate
  \[ R^*_{\mathrm{SCOMP}} \geq R^*_{\mathrm{DD}} \geq \min \left\{ \frac{1}{\ee \ln 2},\, \frac{1}{\ee \ln 2} \frac{1 - \theta}{\theta} \right\} . \]
In particular, SCOMP is optimal for $\theta \geq 1/2$.
\end{theorem}

\begin{IEEEproof}
Following Lemma \ref{lem}, we have wo show that SCOMP satisfies Property $P$.

As explained above, the SCOMP algorithm begins with the DD algorithm, then takes further steps to ensure a satisfying set is reached. However, if DD already provides a satisfying set, no further steps are taken, and SCOMP halts and outputs the DD set. Hence Property $P$ is satisfied.
\end{IEEEproof}

\subsection{Linear programming approaches} \label{sec:lp}

LP algorithms are based on the fact choosing the smallest satisfying set as an estimate is known to be optimal \cite{aldridge}, but is likely to be impractical, as it requires solving an instance of the NP-hard set cover problem. Specifically, the smallest satisfying set is related to the solution to the $0$--$1$ linear program
  \begin{align*} \text{minimize } \ &\sum_{i=1}^n z_i \\
    \qquad \text{subject to } \  &\sum_{i=1}^n x_{ti} z_i \geq 1 \quad \text{when } y_t = 1, \\
                                   &\sum_{i=1}^n x_{ti} z_i = 0 \quad \text{when } y_t = 0, \\
                                   &z_i \in \{0,1\} . \end{align*}
(Recall that $x_{ti}$ indicates if item $i$ is in test $t$, and $y_t$ is the outcome of test $t$.)
The set $\hat\K =  \{i : z_i = 1 \}$ is the smallest satisfying set.
                                   
LP approaches attempt to estimate the defective set via the relaxed version of the $0$--$1$ problem, 
where the $z_i$s can be any non-negative real numbers. Linear programs like this can be solved quickly using, for example, the simplex algorithm.

There are various heuristics for how to turn the optimal solution $(z_i)$ to the relaxed program into an estimate of the defective set. For the purposes of Theorem \ref{lpthm}, it suffices that in the event that all $z_i$s are $0$ or $1$ then the heuristic chooses $\hat{\mathcal K} = \{i : z_i = 1 \}$ (as any sensible heuristic must).

For example, one could consider the following very crude method: if there is any $i$ with $z_i \neq 0, 1$, declare a global error; else estimate $\hat{\mathcal K} = \{i : z_i = 1 \}$ to be the defective set. Malioutov and Malyutov \cite{mal} suggest using the estimate $\hat{\mathcal K} = \{i : z_i > 0 \}$, and show strong performance on simulated problems. (This was the `Malioutov rule' used for the simulations in Figure \ref{fig:sim}.)
Another suggestion could be to estimate $\hat{\mathcal K} = \{i : z_i \geq 1/2 \}$, or to place each item $i$ in $\hat{\mathcal K}$ independently with probability $z_i$.

Along these lines, the `LiPo' algorithm of Chan et al \cite{chan2} assumes the detector knows $k$ exactly, so the LP relaxation can be instead phrased as a feasibility problem. They show that LiPo achieves a nonzero rate for all $\alpha \in (0,1)$, and specifically
\[ R^*_{\mathrm{LiPo}} \geq \frac{1}{\frac83 \, \text{e}^2 \ln 2} \, \frac{1-\theta}{1+\theta} \approx 0.0732 \, \frac{1-\theta}{1+\theta} . \]

However, by using Property $P$, we can show a higher rate, which is optimal for $\theta > 1/2$.

\begin{theorem} \label{lpthm}
Any LP approach as described above with Bernoulli tests can achieve the rate
  \[ R_{\mathrm{LP}}^* \geq R^*_{\mathrm{DD}} \geq \min \left\{ \frac{1}{\ee \ln 2},\, \frac{1}{\ee \ln 2} \frac{1 - \theta}{\theta} \right\} . \]
In particular, LP approaches are optimal for $\theta \geq 1/2$.
\end{theorem}

\newpage

\begin{IEEEproof}
Following Lemma \ref{lem}, we must show that Property $P$ is satisfied.

Note that any item $i$ in a negative test (a definite nondefective) will have $z_i = 0$ to satisfy the second constraint of the linear program. Further, if a test $t$ contains only one possible (and thus definite) defective $i$, the LP solution must have $z_i \geq 1$ to  satisfy the corresponding constraint, and will choose $z_i = 1$ to minimise $\sum_i z_i$. Finally, if these definite defectives form a satisfying set, all constraints are satisfied, and the LP will set all other $z_i$s to $0$, to minimise $\sum_i z_i$. Hence Property $P$ holds, and we are done.
\end{IEEEproof}

\section*{Acknowledgments}

The author thanks Jonathan Scarlett for useful discussions.

\bibliographystyle{IEEEtran}
\bibliography{IEEEabrv,bibliography}

\end{document}